\documentclass[11pt]{article}
\usepackage{moriond,epsfig}

\def\Journal#1#2#3#4{{#1} {\bf #2}, #3 (#4)}

\def\be{\begin{equation}}
\def\ee{\end{equation}}
\def\bea{\begin{eqnarray}}
\def\eea{\end{eqnarray}}

\begin{document}
\vspace*{4cm}
\title{Models and possible progenitors of gamma-ray bursts at the test field of the observations}

\author{B. Gendre}

\address{ARTEMIS (OCA/CNRS) \& IRAP (OMP/CNRS)}

\author{G. Stratta}

\address{Observatory of Rome}

\author{on behalf of the FIGARO collaboration}

\maketitle\abstracts{
During the last 15 years, a standard paradigm has emerged to explain both the progenitor nature and the observed radiations of gamma-ray bursts. In this work we show three GRBs for which the standard paradigm could be tested with high statistics due to their exceptional spectral and temporal coverage. While GRB 1110205 represents a very good example of the standard scenario, GRB 090102 and GRB 111209A do not fit into the standard paradigm.
}

\section{Introduction}

With the discovery that Gamma-Ray Bursts (GRBs) are cosmological events \cite{ghe12}, a common picture has emerged to explain these events. There are two classes of GRBs: short and long events, separated by the canonical duration value of $T_{90} = 2$ s \cite{kou93}. Short events are thought to originate from the merging of a binary system of compact objects \cite{eic89}, even if some magnetar formation models can also explain the observations \cite{uso92}. Long events are associated with the death of a certain kind of Wolf-Rayet stars in a cataclysmic collapse of the core, the collapsar model \cite{woo86}. 
Several pieces of evidence have confirmed the collapsar model: observation of stellar winds around the progenitors \cite{gen04,gen06}, association of type Ib/c supernovae \cite{hjo03},...

\begin{figure}[h]
\begin{center}
\psfig{figure=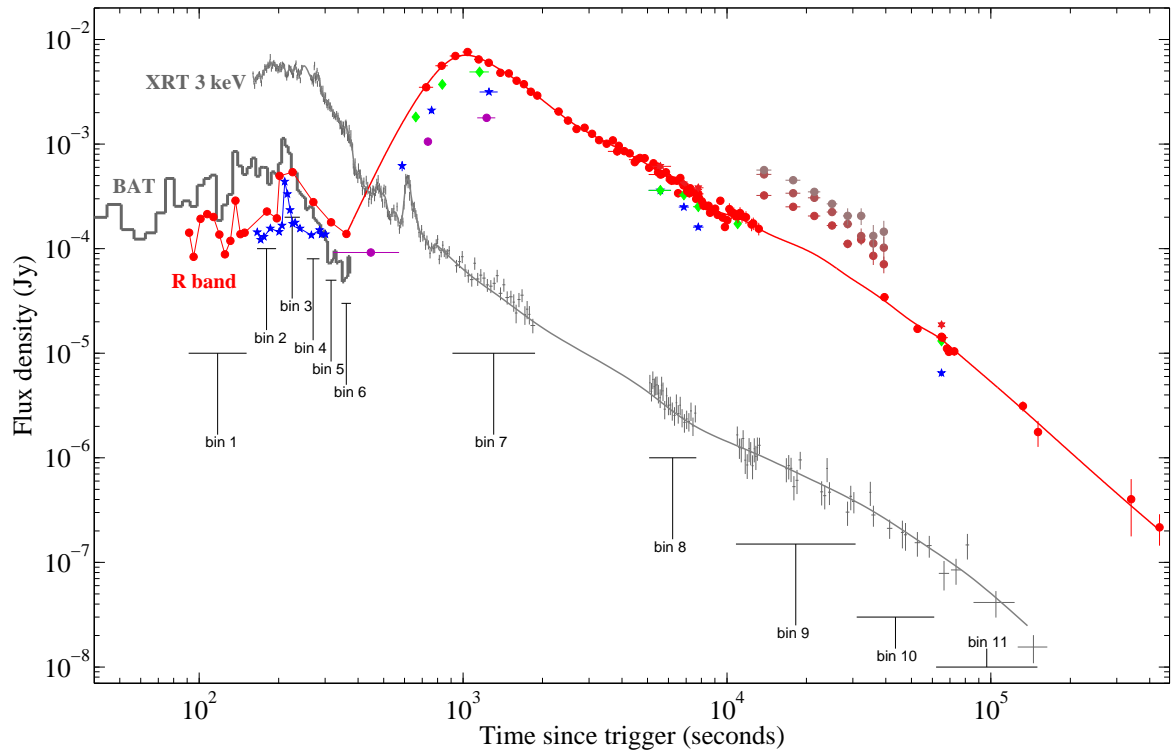,height=100mm}
\caption{Optical light curve of GRB 110205A, extracted from Gendre {\it et al.} 2012. All the components of the fireball model are clearly seen. \label{fig1}}
\end{center}
\end{figure}

Independently on the progenitor nature, a fireball, collimated within a jet and with a Lorentz factor of the order of several hundreds, is produced \cite{ree92,mes97,pan98}. The fireball is responsible of the observed radiation  through three different classes of shocks: internal shocks (the origin of the prompt phase), external shocks (responsible of the afterglow), and the reverse shock (that cause a rebrightening at low energy) as explained in details in the review of \cite{pir05}. A good example can be seen in Fig. \ref{fig1} where the light curve of GRB 110205A is shown. This burst is the archetypal event that fit the fireball model, displaying all the components well separated in time. More details on this burst can be found in Gendre {\it et al.} 2012 \cite{gen11}.


\section{An example of "non-fitting" burst: GRB 090102}

One of the best example of burst that cannot be fitted by the fireball model is GRB 090102. This work has been presented in Gendre {\it et al.} 2010 \cite{gen10}, and can be summarized as follow. In X-rays, it presents a very smooth light curve with no hint of a temporal break. In the optical, the light curve presents a steep-flat behavior, with a break time at $\sim$ 1 ks after the burst. When taken alone, each of the observation band results can be explained by the standard model. In X-ray, this is a typical afterglow expanding in the interstellar medium. In optical, the data could be interpreted as either due to a termination shock, locating the end of the free-wind bubble at the position of the optical break; or as a normal fireball expanding in an ISM, with a reverse shock present at an early time (before the break time). However, once combined together, the flux levels are not compatible between the optical and X-ray band. The cannonball model \cite{dad09} can partly reproduce the data. It appears clear, however, that in order to explain the broad-band emission, some fine-tuning of this model is mandatory, likewise for the fireball model. Another very good example of "non-fitting" burst is GRB 061126 that also feature an unusual afterglow, and again a non standard model has been proposed for explaining this event \cite{per08}.

\section{Peculiar progenitor: the case of GRB 111209A}

\begin{figure}[h]
\begin{center}
\psfig{figure=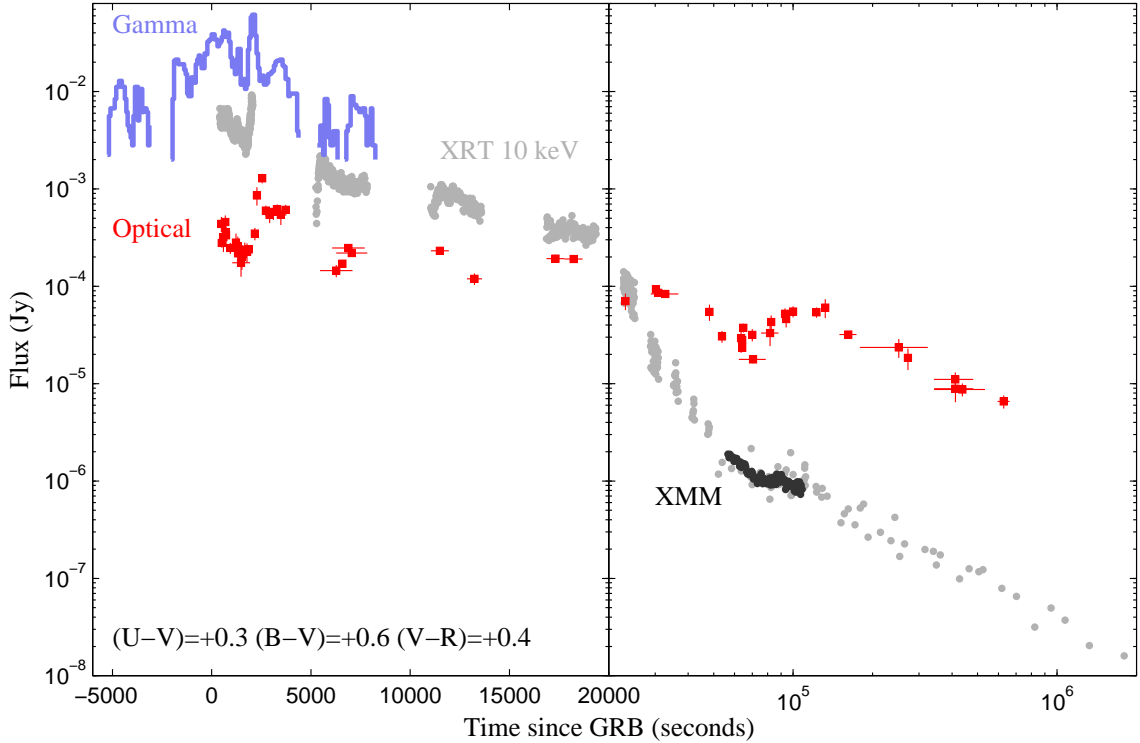,height=100mm}
\caption{Light curve of GRB 111209A. We present the high energy, X-ray and optical light curves. \label{fig2}}
\end{center}
\end{figure}

We now turn our focus to GRB 111209A, a very peculiar event. It was discovered by the Swift satellite, producing two triggers of the Burst Alert Telescope, and also followed by Konus-Wind. As shown by a re-analysis on ground, the burst started about 5400 seconds before $T_0$, and lasted in gamma-ray about 15 000 seconds. In X-ray, the start of the steep decline is supposed to be the true end of the prompt phase \cite{zha06}: taking this stop time we reached a total duration for this event of more than 25 000 seconds \cite{gen12}. Browsing all available archives and catalog of GRBs, it was impossible to find another burst with such a large duration. Because this burst occurred at a redshift of \emph{z} = 0.677 \cite{vre11}, its intrinsic duration is larger than 10 000 seconds, making it the first ultra-long burst studied. Its light curve is displayed in Fig. \ref{fig2}.

\begin{figure}[h]
\begin{center}
\psfig{figure=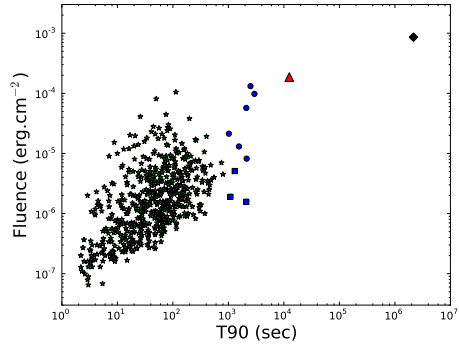,height=50mm}
\caption{Position of GRB 111209A (red triangle) in the fluence-duration plane, compared to normal long GRBs (green stars), tidal disruption events (black diamond), supernovae shock breakouts (blue square), and unknown very long events (blue circles).\label{fig3}}
\end{center}
\end{figure}

The origin of this event is not clear. As one can clearly see in Fig. \ref{fig3}, it is very different from normal long GRBs. It presents a thermal component at the start of the XRT observation. Two super long GRBs (GRB 060218, z = 0.033, and GRB 100316D, z = 0.059) were associated with a supernova shock breakout \cite{cam06,sta11}, and presented a strong thermal component. However the thermal component for GRB 111209A disappears very soon, and most of the prompt phase is free of thermal emission. Moreover, there is no clear evidence of any SN emission. We can therefore discard this hypothesis, and test unusual progenitors. 

The main difficulty in explaining the nature of the progenitor of GRB 111209A is its duration. 
In \cite{gen12} we discuss how a magnetar model cannot reproduce the energetics and spectral characteristics. 
The most probable scenario is a single supergiant star with low metallicity. The hypothesized progenitors of long duration GRBs are Wolf-Rayet stars (stars with the outer layers expelled during stellar evolution). When these layers are still present, as in low metallicity super-giant stars with weak stellar winds, the stellar envelope may fall-back and accretion can fuel the central engine for a much longer time. In this scenario, blue super-giant stars can produce GRBs with prompt emission lasting about $10^4$ seconds \cite{woo12}.

The afterglow analysis uses the observations of XMM-Newton and Swift in the X-ray, TAROT and GROND in the optical, and ACTA data in radio. From the optical-to-gamma-ray prompt spectral energy distribution we find evidence of dust extinction of the order of $A_V\sim0.3-1.5$ mag in the rest frame of the GRB, depending on the assumed spectral continuum, that however is not confirmed during the afterglow emission. 
We find that our results point against a low metallicity environment possibly challenging the low-metallicity progenitor solution. Despite the unusual progenitor nature, the standard fireball model can fit the afterglow data.

\section{Conclusion}

We have presented several cases that does not follow the standard paradigms of GRBs. These cases show that not all GRBs can be explained using the standard fireball, and that not all long GRBs are due to the same kind of progenitor. We have also shown that an unusual progenitor does not imply an unusual afterglow. The exact explanation of GRBs is still not clearly set, and further works are still needed.

\section*{Acknowledgments}

The FIGARO collaboration is funded by the Programme National des Hautes Energies in France.


\section*{References}
\begin{thebibliography}{99}

\bibitem{ghe12} N. Gehrels \& P. Meszaros, \Journal{Science}{337}{932}{2012}.


\bibitem{kou93} C. Kouveliotou, C.A. Meegan, G.J. Fishman, {\it et al.}, \Journal{ApJ}{413}{L101}{1993}.

\bibitem{eic89} D. Eichler, {\it et al.}, \Journal{Nature}{340}{126}{1989}.

\bibitem{uso92} V.V. Usov, \Journal{Nature}{389}{635}{1992}.

\bibitem{woo86} S.E. Woosley, \Journal{ApJ}{405}{273}{1993}.

\bibitem{gen04} B. Gendre, L. Piro, \& M. DePasquale, \Journal{A\&A}{424}{L27}{2004}.

\bibitem{gen06} B. Gendre, A. Galli, A. Corsi, {\it et al.} \Journal{A\&A}{462}{565}{2007}.

\bibitem{hjo03} J. Hjorth {\it et al.}, \Journal{Nature}{423}{847}{2003}.

\bibitem{ree92} M.J. Rees \& P. M\'esz\'aros, \Journal{MNRAS}{258}{41}{1992}.

\bibitem{mes97} P. M\'esz\'aros, \& M.J. Rees, \Journal{ApJ}{476}{232}{1997}.

\bibitem{pan98} A. Panaitescu, P. M\'esz\'aros, \& M.J. Rees, \Journal{ApJ}{503}{314}{1998}.

\bibitem{pir05} T. Piran, \Journal{RvMP}{76}{1143}{2005}.

\bibitem{gen11} B. Gendre, J.L. Atteia, M. Bo\"er, {\it et al.}, \Journal{ApJ}{748}{59}{2012}.

\bibitem{gen10} B. Gendre, A. Klotz, E. Palazzi, {\it et al.}, \Journal{MNRAS}{405}{2372}{2010}.

\bibitem{dad09} S. Dado, A. Dar, \& A. De R\'ujula, \Journal{ApJ}{696}{994}{2009}.

\bibitem{per08} D.A. Perley, J.S. Bloom, N.R. Butler, {\it et al.}, \Journal{ApJ}{672}{449}{2008}.

\bibitem{zha06} B. Zhang, Y.Z. Fan, J. Dyks, {\it et al.}, \Journal{ApJ}{642}{354}{2006}.

\bibitem{gen12} B. Gendre, G. Stratta, J.L. Atteia, {\it et al.}, \Journal{ApJ}{766}{30}{2013}.

\bibitem{vre11} P. Vreeswijk, J. Fynbo, A. Melandri, \Journal{GCN}{12648}{1}{2011}

\bibitem{cam06} S. Campana, V. Mangano, A.J. Blustin, {\it et al.}, \Journal{Nature}{442}{1008}{2006}

\bibitem{sta11} R.L.C. Starling, K. Wiersema, A.J. Levan, {\it et al.}, \Journal{MNRAS}{411}{2792}{2011}

\bibitem{met11} B.D. Metzger, D. Giannios, T.A. Thompson, N. Bucciantini, \& E. Quataert, \Journal{MNRAS}{413}{2031}{2011}

\bibitem{woo12} S.E. Woosley, A. Heger, \Journal{ApJ}{752}{32}{2012}









\end{thebibliography}
\end{document}